\documentclass[12pt]{article}

\usepackage[plain,noend]{algorithm2e}
\usepackage{algorithmic, color}

\usepackage{amsmath,amssymb}
\usepackage{subcaption}
\usepackage{graphicx}
\usepackage{caption}
\usepackage{color}
\usepackage{dcolumn}
\usepackage{bm}
\usepackage{float}
\usepackage{hyperref} 
\usepackage{apacite}
\hypersetup{ colorlinks = true, citecolor = black, urlcolor = blue}

\definecolor{background-color}{gray}{0.98}

\title{A review of Approximate Bayesian Computation methods via density estimation: inference for simulator-models}

\author{Clara Grazian and Yanan Fan\thanks{School of Mathematics and Statistics, UNSW Sydney, Kensington 2052, Australia}}

\date{}
\begin{document}
\maketitle

\begin{center}
\subsubsection*{\small Article Type:}
Advanced Review

\hfill \break
\thanks

\subsubsection*{Abstract}
\begin{flushleft}
This paper provides a review of Approximate Bayesian Computation (ABC) methods for carrying out Bayesian posterior inference, through the lens of density estimation. We describe several recent algorithms and make connection with traditional approaches. We show advantages and limitations of models based on parametric approaches and we then draw attention to developments in machine learning, which we believe have the potential to make ABC scalable to higher dimensions and may be the future direction for research in this area. 

\end{flushleft}
\end{center}

\clearpage

\renewcommand{\baselinestretch}{1.5}
\normalsize

\clearpage

\section{Introduction}

The goal of statistical inference is to draw conclusions about properties of a population given finite observational data $\mathbf{y}_0=(y_0^{(1)},\dots,y_0^{(n)})$. This typically proceeds by first specifying a parametric {\it statistical} model for the data generating mechanism $p (\mathbf{y} | \theta)$, where $\theta$ denotes the parameters of the {\it parametric} model. A likelihood function can then be specified based on the parametric form.  Once the data have been observed, the formal Bayesian inferential framework then allows us to combine the likelihood function with any prior information, allowing inference to be carried out based on model parameters $\theta$.

In practice, and particularly when one wants to define realistic models for modern applications, the parametric model can be difficult to specify, and the likelihood function may not always be available in closed form.  Approximate Bayesian Computation (ABC) is a class of tools and algorithms which have been developed to perform Bayesian inference in the absence of a likelihood function. A defining feature of this class of algorithms is the existence and reliance on a known data generating mechanism, so that for any value of $\theta$, we can obtain pseudo-observations using the same mechanism that generated the observed data; we call this {\it simulator-based models}, i.e. models which are specified only through the generative mechanism. This type of modelling has been proposed in several contexts,  examples include  Astrophysics (simulating the formation of galaxies,  \shortciteA{cameron2012approximate}), Neuroscience (simulating neural circuits, \shortciteA{lueckmann17}), Econometrics (\shortciteA{gourieroux1993indirect}), Epidemiology (simulating the spread of bacterial infections, \shortciteA{luciani2009epidemiological}), Ecology and Genetics (simulating animal populations dynamics, \shortciteA{beaumont2010approximate}) and so on. 

More formally, suppose we have a set of $n$ observed data points $\mathbf{y}_0=(y_0^{(1)},\dots,y_0^{(n)})$. The data-generating process is known, but the likelihood function is unavailable, due to the fact that either it is too costly to evaluate or simply cannot be analytically computed. The latter can happen, for example, in likelihoods involving complex integrals, e.g. in population genetics, where there is an integration over coalescence trees \shortcite{cornuet2008inferring} or in stochastic volatility models where there is an integration over all the observational period of time \shortcite{creel2015abc}. Then, given a particular parameter value (which may be simulated from its prior distribution), we simulate a new set of pseudo-observations $\mathbf{y}_{sim}$ of the same dimension 
\begin{equation*}
\mathbf{y}_{sim} \sim p(\mathbf{y} | \theta)
\end{equation*}
where we use the same notation to denote the true but unknown parametric model. Here we no longer have the analytical form of the probability distribution function (pdf), and instead we are able to obtain pseudo-observations $\mathbf{y}_{sim}$. 

The likelihood function is approximated via simulations of the parameter and data pair $(\theta, \mathbf{y}_{sim})$, instead of being analytically evaluated at $\mathbf{y}_0$. Earlier works on ABC were inextricably intertwined with algorithms, from the rejection-ABC algorithms of \shortciteA{tavare97} and \shortciteA{pritchard1999population}, to MCMC-ABC algorithms (\shortciteA{marjoram03} and \shortciteA{bortot07}), to the more advanced use of sequential Monte Carlo algorithms (\shortciteA{sisson2007sequential} and \shortciteA{beaumont2010approximate}). \shortciteA{sissonfan2018} provides a review on these types of algorithmic approaches. 

The mainstream approach in ABC is
 to compare the observed data with the simulated data,  usually through the use of some discrepancy measure. If the two are within some small distance of each other, then the parameter value $\theta$ that generated the pseudo-observations would be kept and constitutes as a sample from the posterior distribution. In order to increase the efficiency of the algorithms, the discrepancy measure is applied on summary statistics $\eta(\mathbf{y})$ of the data $\mathbf{y}$ instead of the full set of data. Therefore, the parameter is studied through a likelihood function conditional on observed summary statistics, instead of data \shortcite{wilkinson2013approximate} and consequently the target likelihood function is
 \begin{equation}\label{approxL}
L(\theta)\approx p(\eta(\mathbf{y}_0)|\theta),
\end{equation}
for some choice of summary statistics $\eta(\cdot)$. In doing so, there may be a loss of information unless the summary statistics are sufficient (which is unlikely since traditional approaches to finding sufficient statistics require some knowledge of the likelihood function). Details on the selection of summary statistics and discrepancy measures and related aspects of inference based on this approach are summarised in \shortciteA{sisson+fb2018}.
 
 Nevertheless, inference based on the above approach can be slow and highly inefficient and can become computationally intractable when the dimension of the parameter $\theta$ is large, as many more datasets will need to be simulated to obtain a good approximation.   

 In this work, we focus our review on inference based on approximations of the likelihood, that is, methods that attempt to directly approximate the likelihood $L(\theta)$. Such methods have the advantage that they can be considerably more efficient because they no longer depend on minimising a discrepancy measure and in some cases, do not even require the elicitation of summary statistics. We begin by reviewing approximation methods where some parametric form is assumed for the unknown likelihood in Section \ref{sec:syntlik}; in Section \ref{sec:nonparam} we describe nonparametric approaches to approximating the likelihood functions, and make some connections with the standard ABC approaches; in Section \ref{sec:machine}, we describe some recent developments using machine learning methods which offer the potential of scaling up to higher dimensions; we provide some examples to illustrate the methods described in this review in Section \ref{sec:examples}, and conclude with some discussion in Section \ref{sec:discussion}. 
 In what follows, we use the simplified notations $\eta(\mathbf{y}_0)=\eta_0$ and $\eta(\mathbf{y}_{sim})=\eta_{sim}$. 

\section{Parametric likelihood}
\label{sec:syntlik}

If the model for the data or summary statistics can be considered reasonably regular (for instance, if the summary statistic is some type of sample average and if the sample size is large) then it may be reasonable to approximate the distribution of the summary statistics by a Gaussian:
\begin{equation}
p(\eta_0 | \theta) = \frac{1}{(2 \pi) ^{p/2}\mid \Sigma_{\theta}\mid ^{1/2} } \exp \left\{ -\frac{1}{2} \left[ (\eta_0 - \mu_{\theta})^T \Sigma_{\theta}^{-1} (\eta_0 - \mu_{\theta})\right] \right\},
\label{eq:synthlik}
\end{equation}
where the expected value $\mu_{\theta}$ and the variance  $\Sigma_{\theta}$ are, in general, unknown and $p$ is the dimension of the summary statistics. Equation \eqref{eq:synthlik} represents an approximation of the likelihood \eqref{approxL}, unless the distribution of the summary statistics is indeed Gaussian. The goodness of the approximation depends on the validity of the asymptotic normality assumption.
The parameters of the Gaussian are approximated by averages over a set of simulated summary statistics:
\begin{align*}
\hat{\mu}_{\theta} &= \frac{1}{N} \sum_{j=1}^N \eta_{sim}^{(j)} \\
\hat{\Sigma}_{\theta} &= \frac{1}{N-1}\sum_{j=1}^N (\eta_{sim}^{(j)} - \hat{\mu}_{\theta})(\eta_{sim}^{(j)} - \hat{\mu}_{\theta})^T,
\end{align*}
where $N$ is the total number of datasets simulated at $\theta$. Thus, $N$ is a trade-off parameter: as it increases, the approximation becomes more accurate, at the expense of an increased computational burden. This approach produces an approximation of the likelihood, $\hat{L}_s$, termed synthetic likelihood by \citeA{wood2010statistical}.

Inference for $\theta$ can be obtained by directly maximising the synthetic likelihood or by deriving the posterior distribution in a standard Bayesian context,  using the approximate likelihood of the summary statistics as the likelihood and a standard algorithm such as the Metropolis-Hastings algorithm to traverse through the space of $\theta$ \shortcite{meeds2014gps}. Note that the approximated Gaussian distribution $\mathcal{N}(\hat{\mu}_{\theta}, \hat{\Sigma}_{\theta})$ is not an unbiased estimate of the synthetic likelihood $\mathcal{N}(\mu_{\theta}, \Sigma_{\theta})$. \shortciteA{price2018bayesian} analize the use of the unbiased estimator proposed by \shortciteA{ghurye1969unbiased}, and that the posterior distribution is of the form: 
\begin{equation*}
\pi(\theta | \mu_{\theta}, \Sigma_{\theta}) \propto \mathcal{N}(\eta_0 ; \mu_{\theta}. \Sigma_{\theta}) \pi(\theta).
\end{equation*}
Under the assumption that the summary statistics are Gaussian distributed, the synthetic likelihood allows us to target the posterior distribution $\pi(\theta | \mu_{\theta}, \Sigma_{\theta})$ but not directly $\pi(\theta | \eta_0)$, particularly if a transformation of the summary statistics is used to encourage normality. 

The use of the synthetic likelihood does not require the user to define a discrepancy function between summary statistics, since it is implicitly induced by the use of a multivariate Gaussian distribution. However, empirical work shows that the synthetic likelihood can be robust to the violation of normality, as shown in \shortciteA{everitt2017bootstrapped} and \shortciteA{price2018bayesian}. The quality of the approximation depends on how close the distribution of the summary statistics is to a Gaussian distribution. In particular, for small sample size the approximation may be unreliable, see Example \ref{sub:ricker}.  Recently, a more robust semiparametric version has been proposed to relax the normality assumptions \shortcite{an2018robust}.

As described above, the synthetic likelihood can be used as a surrogate for the likelihood in a Bayesian approach based on standard MCMC algorithms or can be integrated within an ABC framework \shortcite{meeds2014gps}, by convoluting a kernel measuring the discrepancy between observed and simulated datasets with respect to the Gaussian synthetic likelihood:

\begin{equation*}
L_s(\theta) = \int \mathcal{K}_{\varepsilon} (\eta_0,\eta_{sim}) \mathcal{N}(\eta_{sim};  \hat{\mu}_{\theta}, \hat{\Sigma}_{\theta})d\eta_{sim}.
\end{equation*}
Here the simulated datasets/summary statistics are assumed to be Gaussian distributed, however it is possible to introduce an additional error term allowing for a different model for the observed summary statistics. If a Gaussian kernel is used, $$\mathcal{K}_{\varepsilon}(\eta_0,\eta_{sim}) = \frac{1}{(2\pi \varepsilon)^{n/2}}\exp(-\frac{1}{2\varepsilon^2} (\eta_{sim}-\eta_0)^T (\eta_{sim}-\eta_0)),$$
then the approximated likelihood function $\hat{L}_s$ is again a Gaussian, $L_s(\theta) = \mathcal{N}(\hat{\mu}_{\theta} , \hat{\Sigma}_{\theta} + \varepsilon^2 \mathcal{I})$ where $\mathcal{I}$ is the identity matrix. By allowing $\varepsilon \rightarrow 0$, the bias introduced by using simulated datasets can be reduced. In the latter procedure,
the likelihood function depends on a kernel which is a function of the difference between summary statistics, and it is more robust for irregular models, such as those in chaotic dynamic systems, where the Gaussian assumption can be seen as too strong. An alternative approach would be estimating the covariance matrix in a robust way, as proposed in \shortciteA{wood2010statistical}. 

Under this synthetic likelihood approach, the step simulating new datasets can be particularly expensive; various approaches to work more efficiently with the synthetic likelihood has been proposed, see, for example, \shortciteA{meeds2014gps}.

\section{Nonparametric likelihoods}
\label{sec:nonparam}

The vanilla Approximate Bayesian Computational algorithm uses a discrepancy measure to compute the distance among summary statistics between simulated and observed datasets. Algorithm \ref{alg1} below describes a standard rejection sampling ABC algorithm, see also \shortciteA{pritchard1999population}. 

\begin{algorithm}[h]
\caption{Rejection-ABC: Discrepancy measure $\Delta$ is usually taken to be the Euclidean distance; tolerance level $\varepsilon$ should be set close to 0 (the particular choice is problem-specific).}
\label{alg1}
    \begin{algorithmic} 
        \REQUIRE 
        1. Observed data $\mathbf{y}_0$; 2. Prior distribution $\pi(\theta)$; 3. Generative model $p(\mathbf{y}|\theta)$; \\ 4. Discrepancy measure $\Delta$; 5. tolerance value $\varepsilon$:
        \STATE 1. Simulate $\theta^{*} \sim \pi(\theta)$
        \STATE 2. Simulate $\mathbf{y}_{sim} \sim p(\mathbf{y} | \theta^{*})$
        \IF{$\mathbf{y}_{sim} = \mathbf{y}_0$ (discrete data) or $\Delta(\eta_0, \eta_{sim})< \varepsilon $ (continuous data)}
            \STATE $\theta^{*}$ forms a part of the posterior sample;
        \ELSE
            \STATE Discard $\theta^{*}$.
        \ENDIF
        \STATE 3. Repeat Steps 1 and 2 until enough samples of $\theta$ are obtained.
        \vspace{0.3cm}
    \end{algorithmic}
\end{algorithm} 

In this basic version of the algorithm, when the data is discrete, it is possible to consider matching the simulated data with the observed data perfectly, however this approach is not possible when the data is continuous and it is also highly inefficient as the sample size increases.  Therefore, in practice, matching within a small distance of $\varepsilon \geq 0$ is considered close enough. If $\varepsilon$ is too large, posterior estimates of $\theta$ can be biased, and posterior credible intervals will be too large. As $\varepsilon$ decreases, computational cost increases dramatically, so there is a trade-off between accuracy and computational capacity. The use of low-dimensional summary statistics is necessary in most applications, despite the fact that their use introduces an additional layer of complexity and potential loss of information from the data. However, well chosen summary statistics can enhance inference \shortcite{fan+ns13}.  

The most used modification to improve computational efficiency of Algorithm \ref{alg1} is the so-called regression adjustment \shortcite{beaumont2002approximate}. The main idea of regression ABC is running standard ABC with a relatively large threshold level $\varepsilon$ and then adjust the obtained samples through a regression
\begin{equation*}
\label{eq:regmod}
\theta_j = f(\eta_j) + \gamma_j
\end{equation*}
where $j=1,\dots,J$, $J$ is the total number of values accepted after running Algorithm \ref{alg1} and $\gamma_j$ is an error term centred around zero. The resulting values then become closer to samples from the posterior distribution as the regression predicts $\theta$ at $\varepsilon=0$. It can be proved that the empirical variance of the adjusted sample is smaller than the empirical variance of the non-adjusted values \shortcite{blum2010approximateblum}. More recently, \shortciteA{li2016improved} proved that, for an appropriate choice of the bandwidth in ABC, standard ABC and regression-adjusted methods lead to an approximate posterior distribution which, asymptotically, correctly quantifies the uncertainty and, in particular, the threshold $\varepsilon$ is required to depends on the sample size $n$ in order to achieve such result.

Even with adjustments for the output of standard ABC, most parameter values can produce large distances between summary statistics when simulation is performed from an uninformative prior distribution, so a large number of simulated datasets is needed to identify the area of the parameter space closest to the true parameter value, and this is computationally costly. An alternative approach would be to model the discrepancy measure.
\shortciteA{gutmann2016bayesian} showed that the likelihood function can be approximated by
\begin{equation*}
p(\eta_0 
| \theta) \approx \mathbb{E}[\mathcal{K}(\eta_0,\eta_{sim})],
\end{equation*}
where $\mathcal{K}(\cdot)$ is a kernel function evaluating the distance among observed and simulated summary statistics.
This expected value can then be approximated by
\begin{equation*}
\hat{L}_{\mathcal{K}}(\theta) = \frac{1}{N} \sum_{j=1}^N \mathcal{K}(\eta_0,\eta_{sim}^{(j)}),
\end{equation*}
that is, we can repeatedly sample $\eta_{sim}$ at a given value of $\theta$, and use a kernel density estimator to approximate the likelihood at $\theta$. The connection between traditional ABC algorithms and kernel density estimation was also explored in \shortciteA{sisson+fb2018} and \shortciteA{blum2010approximateblum} for an earlier reference. 

Alternatively, the approximate likelihood function $\hat{L}_{\mathcal{K}}$ can be written as a function of the discrepancy, $\Delta$ in Algorithm \ref{alg1}. The most used function is the Uniform kernel, for which $\mathcal{K}(\Delta) = c \mathbb{I}_{[0,\varepsilon)}(\Delta)$, where $\mathbb{I}_{[a,b)}(x)$ is an indicator function which is equal to one if $x \in [a,b)$ and $c$ is some constant. Then the likelihood function can be approximated by 
\begin{equation*}
\hat{L}_{\mathcal{K}}(\theta)  = \hat{\Pr}(\Delta(\eta_0, \eta_{sim}) \leq \varepsilon),
\end{equation*}
i.e. the empirical probability that the discrepancy measure is smaller than a threshold $\varepsilon$. 
\shortciteA{gutmann2016bayesian} showed that maximising the synthetic likelihood (Section \ref{sec:syntlik}) corresponds to maximising a lower bound of a nonparametric approximation of the likelihood $L_{\mathcal{K}}(\theta)$. 

\shortciteA{gutmann2016bayesian} models the discrepancy $\Delta(\eta_0, \eta_{sim})$ as a Gaussian process, using a squared exponential covariance function and uncorrelated Gaussian noise, within the framework of a Bayesian optimization algorithm \shortcite{williams2006gaussian}. The resulting algorithm is defined as Bayesian optimization for likelihood-free inference (BOLFI). For an application in population genetics see \shortciteA{numminen2016impact}. Since the discrepancy is a positive function, it is possible to consider a transformation,  $g(\cdot): \mathbb{R^+ \rightarrow \mathbb{R}}$, of it which will be more likely to follow a Gaussian distribution, for example the logarithmic transform, so that $\Pr(g[\Delta(\eta_0, \eta_{sim})] \leq \varepsilon^{\prime})$. 
However, the Gaussian assumption may not hold in general, in particular, because the variance of the discrepancy is likely to vary over the parameter space; therefore, the Gaussian process model for the discrepancy may affect the accuracy of the ABC posterior estimation when some of these assumptions are not met.  

\shortciteA{jarvenpaa2018gaussian} proposed three methods using a Gaussian process to model the discrepancy measure:
\begin{itemize}
	\item The discrepancy measure $\Delta$ can be assumed to follow a Gaussian distribution $\Delta(\eta_0, \eta_{sim}) \sim \mathcal{N}(f(\theta),\sigma^2)$, i.e. with constant variance. The mean is modelled as a Gaussian process of mean $m(\theta)$ and covariance structure $k(\theta,\theta^{\prime})$ given by some function of the distance between values, for example a squared exponential function. In this setting, 
	\begin{equation*}
	Pr(\Delta(\eta_0, \eta_{sim}) \leq \varepsilon) = \Phi \left( \frac{\varepsilon - \mu(\theta)}{\sqrt{v(\theta)+\sigma^2}}\right)
	\end{equation*}
	where $\Phi(\cdot)$ is the cumulative distribution function of a standard Gaussian variable and $\mu(\theta)$ and $v(\theta)$ are the posterior mean and variance of the function $f(\theta)$.
	\item The variance of the discrepancy can be allowed to vary over the parameter space, so that $\Delta(\eta_0, \eta_{sim}) \sim \mathcal{N}(f(\theta), \sigma^2 \exp(g(\theta)))$. In addition to a prior model for the mean function $f(\theta)$ (which can be again assumed to be a Gaussian process), also the variance or a function of it, e.g. $\log(g(\theta))$, needs to be modelled; if it is reasonable to assume that it changes smoothly as a function of $\theta$, a Gaussian process can be again imposed as prior distribution. In this setting, 
	\begin{equation*}
	Pr(\Delta(\eta_0, \eta_{sim}) \leq \varepsilon) = \Phi \left( \frac{\varepsilon - \mu(\theta)}{\sqrt{v(\theta)+\sigma^2 \exp(g(\theta))}}\right).
	\end{equation*}
	\item Instead of directly modelling the discrepancy measure, it is possible to associate it to a latent variable $Z$, such that $Z = 2 \mathbb{I}_{\Delta(\eta_0, \eta_{sim}) \leq \varepsilon} - 1$ can take value in $\{ -1, +1 \}$, following a probit or a logit model: $\Pr(z | f(\theta)) = g(z,f(\theta))$ and where the function $f(\theta) \sim \mathcal{N}(m(\theta), k(\theta,\theta^{\prime}))$. The difference between this version of the algorithm and the two previous ones is that here the discriminative function is modelled as a smooth function, but no assumption on the form of the distribution of the discrepancy measure is made. 
\end{itemize}
An alternative would be to consider a Student't $t$ distribution, as in \shortciteA{shah2014student}. 
The accuracy of the estimation depends on how well a Gaussian distribution can model the distribution of the discrepancy measure: if the discrepancy is roughly Gaussian distributed, the standard GP or the heteroskedastic version can be a good representations, while, as it moves away from normality (with multimodality or heavy tails), the classification approach, the Student's $t$ distribution or other algorithms as the ones presented in Section \ref{sec:machine} can lead to better estimates.
Since the goodness of fit of each of these models depends on the specific problem at hand, it is possible to compare several models through model selection techniques as part of the analysis. \shortciteA{jarvenpaa2018gaussian} proposes two utility functions along the line of \shortciteA{vehtari2012survey}: the mean of the log-predictive density, which measures how well the Gaussian process predicts the distribution of the discrepancies, and the classifier utility, which penalizes realisations of the discrepancy that are under the threshold.

The approaches based on using Gaussian process priors for the discrepancy measure are different from \shortciteA{wilkinson2014accelerating}, where the likelihood is directly modelled as a Gaussian process, or \shortciteA{meeds2014gps} where each element of the intractable mean and covariance matrix is modelled as a Gaussian process. Clearly here the choice of the discrepancy is essential, as it affects the goodness of fit of the Gaussian process and, therefore, the quality of the approximation. The standard choice is the Euclidean distance.

A nonparametric approach (based on the definition of a discrepancy measure between summary statistics) can be more accurate than a parametric approach as in Section \ref{sec:syntlik} when the summary statistics $\eta(\cdot)$ is low-dimensional; however, as the number of summary statistics increases, the accuracy of the algorithms tends to deteriorate. 
The methods just described use either kernel density estimates or Gaussian processes to model the distribution of the discrepancy measure in order to reduce the assumptions on the distribution of the summary statistics. 
A completely nonparametric alternative to these methods within the ABC framework involve making use of an empirical likelihood approach \shortcite{owen1988empirical,mengersen2013bayesian}. The empirical likelihood is a nonparametric estimator of the likelihood function. Given a set of independent and identical distributed observations $y_i$, $i=1,\dots,n$ from a distribution $F$, the empirical likelihood function is defined as a set of weights
\begin{equation*}
L_{el}(p) = \max_{p_i} \prod_{i=1}^n p_i
\end{equation*}
where $p_i$ are obtained under constraints $\sum_{i=1}^n p_i = 1$, $0 < p_i < 1$ and $\sum_{i=1}^n p_i h(y_i,\theta) = 0$. Here, $\sum_{i=1}^n p_i h(y_i,\theta) = 0$ is a moment condition. Extension to non-i.i.d. settings are available in \citeA{owen2001empirical} and  \shortciteA{schennach2005bayesian,grendar2009asymptotic} provide a Bayesian justification of this procedure. 
The empirical likelihood approach defines a set of weights for the values of the parameter of interest. By combining simulations from the prior distribution with the weights defined through the empirical likelihood, it is possible to obtain an approximate sample from the posterior distribution. With respect to standard ABC methods, this approach avoids the definition of summary statistics, whose relationship with the parameters of the model is, in general, unknown. The method can be applied in settings where model miss-specification might lead to strongly biased estimates \shortcite{grazian2017approximate}. However the definition of unbiased estimating equations $h(\cdot)$ is not always straightforward.

\section{Scaling conditional density estimation with mixtures and neural networks}\label{sec:machine}

As has been discussed, standard ABC algorithms often rely on the definition of a similarity threshold $\varepsilon$ and on the approximation of the posterior distribution $\pi(\theta | \mathbf{y})$ with a distribution conditional on the similarity among datasets $\pi(\theta | \Delta(\eta_0,\eta_{sim}) \leq \varepsilon)$. However, this approach presents some drawbacks, since the accuracy increases only when $\varepsilon \rightarrow 0$, at the expense of an increased computational cost. The lack of scalability to higher dimensions is well-recognised in the ABC literature
\shortcite{sissonfb18}, this drives more recent research in the direction towards scalable and more user friendly techniques.

An active stream of current research focuses on treating the likelihood in Equation \ref{approxL} as a regression density estimation problem, using simulations of both the parameters and the summary statistics to train the conditional density.
The resulting estimated density is then an analytically tractable approximation of the likelihood function then used for a Bayesian analysis. Note that directly approximating the likelihood involves an additional step to compute either the posterior or maximising the likelihood to obtain the MLE for $\theta$. Such approaches may be preferable in problems where, for example,  inference is required for 
multiple datasets arising from the same model. 

\shortciteA{fan+ns13} describes a
flexible conditional density estimation approach where the approximation is constructed from a sample of $N$ summary statistic and parameter pairs
$(\eta^1,\theta^1),\dots,(\eta^N,\theta^N)$ drawn from a distribution $p(\eta|\theta)h(\theta)$.  
Note that unlike the standard ABC algorithms such as the rejection-ABC described in Algorithm \ref{alg1}, while the summary statistics are generated given $\theta$ from the sampling distribution for the intractable model of interest, the parameters are not necessarily generated from the prior.  Instead, $h(\theta)$ is a distribution chosen to reflect the region in the vicinity of $\eta_0$, since interest is in the posterior distribution $\pi(\theta |\eta_0)$.  Some rough knowledge of the high likelihood region of the parameter space is needed.
\shortciteA{fan+ns13} suggest initial pilot analysis for setting $h(\theta)$.

Since a good approximation of the conditional distribution is crucial in this approach, two issues in particular should be considered in designing the density estimator. First, the relationship between $\eta$ and $\theta$ can be complex, careful selection of summary statistics may help simplify this relationship. Second, when the dimension of the summary statistic is large, density estimation for the joint distribution is difficult. \shortciteA{fan+ns13} advocated a two step approach.

The first step is to build marginal regression models for each component of $\eta = (\eta_1,\ldots, \eta_K)$ conditional on $\theta$. A mixture of experts model was used for this purpose coupled with a variational method for fitting of the model based on sampled data ($\eta^j, \theta^j, j=1\ldots, N$), where the regression density estimator for each component $\eta_k$ takes the form
$$
f_k(\eta_k|\theta) = \sum_{m=1}^M w_{km}N(\mu_{km}(\theta), \sigma^2_{km}(\theta)),
$$
for some appropriately chosen value $M$, indicating the total number of mixture components to be used, and $w_{km}(\theta)=\frac{\exp(\xi^{km}_0+(\xi^{km})^{T}\theta)}{\sum_{m=1}^M \exp(\xi^{km}_0+(\xi^{km})^{T}\theta)}$, $\mu_{km}(\theta)=\beta_0^{km}+(\beta^{km})^{T}\theta$ and $\log(\sigma^2_{km}(\theta)=\gamma_0^{km}+(\gamma^{km})^{T}\theta$.

Then, a conditional density estimate for the joint distribution of $\eta$ given $\theta$ is constructed.  The data $(\eta^j,\theta^j)$ are transformed to $(U^j,\theta^j)$, where $U_k^j=\Phi^{-1}(\hat{F}_k(\eta_k^j|\theta^j))$, $\hat{F}_k(\eta_k|\theta)$ is the distribution function corresponding to the density $\hat{f}_k(\eta_k|\theta)$.  If the marginal densities for each
$\eta_k$ are well estimated, the transformation to $U^j$ makes each component of $U^j$ approximately standard Gaussian regardless of the value of $\theta$.  
A mixture of Gaussian distributions is then fitted to the data $(U^j,\theta^j)$, $j=1,\dots,N$. The joint density of $(U,\theta)$ is a mixture of multivariate Gaussians taking the form
$$
g(U,\theta)=\sum_{l=1}^L w_lN(\mu_l, \Psi_l)
$$
with $L$ Gaussian mixture components, and $w_l$, $\mu_l$ and $\Psi_l$ are respectively the weight, mean and covariance matrix corresponding to the $l^{th}$ mixture component.

The conditional distribution of $U|\theta$, implied by the multivariate Gaussian mixture of the joint distribution, is again a mixture of Gaussians, denoting this new mixture by
$$
g(U|\theta)=\sum_{l=1}^L w^c_l N(\mu^c_l, \Psi^c_l)
$$ 
where $\mu^c_l$ and $\Psi^c_l$ are the conditional mean and covariance of $U|\theta$ in the $l^{th}$ component of the multivariate $N(\mu_l,\Psi_l)$. The mixing weights for the conditional distribution is given as
$$
w_l^c = \frac{w_l\phi(\theta, \mu_l, \Psi_l)}{\sum_{m=1}^M w_m\phi(\theta, \mu_m, \Psi_m)},
$$
where $\phi(\theta, \mu_l, \Psi_l)$ denotes the multivariate Gaussian density in $\theta$ with mean $\mu_l$ and $\Psi_l$ implied by $g(U,\theta)$. 

Finally, the approximated likelihood of $\eta|\theta$ is derived through back-transformation, via
$$
\hat{L}(\eta|\theta)=\hat{g}(U|\theta)\prod_{k=1}^K\frac{\hat{f}_k(\eta_k|\theta)}{\phi(U_k;, 0,1)}.
$$
The rationale behind the two stage approach is that we can estimate the marginal distributions well without a huge amount of data, and this in turn will improve estimates on the joint distribution, based on a moderate amount of data.

More recent developments have focussed on the use of conditional neural density estimators. In short, a neural density estimator is a parametric model $q_{\phi}$, for example a neural network parameterised by the weights $\phi$. The function $q_{\phi}$ is trained by maximising the total log probability $\sum_{j=1}^N \log q_{\phi}(\mathbf{y}^j|\theta^j)$ with respect to $\phi$. Given sufficient training data and a sufficiently flexible model, $q_{\phi}(\mathbf{y}|\theta)$ will approximate the conditional distribution $p(\mathbf{y}|\theta)$. \shortciteA{papamakarios19} proposed a sequential neural density estimator using conditional autoregressive flows \shortcite{papamakarios17} aimed at a general purpose solution, with an adaptive online scheme for the choice of a sampling distribution $h(\theta)$. Recall that the prior is generally too diffuse to be informative enough to sample $\theta$ in the region corresponding to $\mathbf{y}_0$.

Clearly, any conditional density estimation to approximate the likelihood can be used to approximate the posterior distribution $p(\theta|\mathbf{y}_0)$ directly. However, direct approximation of the posterior distribution is complicated by the need to sample $\theta$ from $h(\theta)$ rather than the prior $\pi(\theta)$. A sequential neural posterior estimator was used, for example, in \shortciteA{papamakarios2016fast}, who used a Mixture Density Network (MDN), i.e. a mixture of $K$ Gaussian distributions whose parameters are estimated by a neural network, while \shortciteA{lueckmann2018likelihood} and \shortciteA{lueckmann17} used multi-layer neural networks. The sampling distribution $h(\theta)$ is obtained over several quick iterations of posterior density estimation, with the improved estimate of $h(\theta)$ being based on the previous estimate of the posterior distribution. 

Finally, a correction is required to account for the fact that the samples of $\theta$ are not drawn from the prior. The conditional density function approximates  
\begin{equation*}
{\tilde q}_{\phi}(\theta | \mathbf{y}) \propto h(\theta) \pi( \mathbf{y}\mid \theta),
\end{equation*}
which is not the true posterior; to obtain an approximation of the posterior distribution, it is necessary to weight it as
\begin{equation}\label{PostAdj}
\hat{\pi}(\theta \mid \mathbf{y}_0) \propto \frac{\pi(\theta)}{h(\theta)} {\tilde q}_{\phi}(\theta | \mathbf{y}_0).
\end{equation}
where ${\tilde q}_{\phi}(\theta \mid \mathbf{y})$ can be a neural network parameterised by $\phi$ and trained on samples of $\theta$ and $\mathbf{y}$.
Since reweighting (Equation \ref{PostAdj}) introduces extra variance into the estimation, \shortciteA{greenberg19} proposed an automatic posterior transform method called ``automatic posterior transformation'' (APT). Let a proposal posterior be defined as
\begin{equation*}
\hat{q}_{\phi} (\theta | \mathbf{y}) \propto q_{\phi}(\theta | \mathbf{y})\frac{h(\theta)}{\pi(\theta)}
\end{equation*}
where $q_{\phi}(\theta | \mathbf{y})$ is an estimate of the true posterior $\pi(\theta| \mathbf{y})$, and training is carried out by minimizing the objective function $-\sum_{j=1}^N \log \hat{q}_{\phi} (\theta^j | \mathbf{y}^i)$ with respect to $\phi$; the procedure recovers both the true posterior and the proposal posteriors.

Other related works include \shortciteA{alsing19} who uses neural density estimators in the context of problems encountered in cosmology; \shortciteA{radev19} showed a convolutional network in an ABC setting can be trained to directly obtain the posterior mean and variance;
\shortciteA{bonassi+yw11} used multivariate Gaussian mixture 
models for the density estimator in the context of statistical genetic models; and \shortciteA{izbicki18} provided a nonparametric density estimation method aimed at high dimensional data.

\section{Examples}\label{sec:examples}

We now compare some of the techniques presented in the previous Sections on two different examples. First, we analyze data simulated from a Gaussian generative model; this is a benchmark, since in this case it is possible to choose a sufficient summary statistic and there is no loss of information in ABC procedures. The second example uses the Ricker model, which is well-known in the ABC literature since its use in the work of \shortciteA{wood2010statistical}. Both examples presented are based on simulations and we repeated the simulations 250 times for each model, in order to study the frequentist behaviours of the procedures analyzed.
All codes used in this Section are available at the website \url{https://github.com/cgrazian/ABC_review}.

\subsection{Gaussian data}

We first consider a simple setting of observations generated from a Gaussian distribution with known variance. Therefore, the simulated data can be simply generated by 
\begin{equation*}
\mathbf{y}_{sim} = \boldsymbol{\theta} + \boldsymbol{\gamma} \qquad \gamma \sim \mathcal{N}(\mathbf{0}, \mathcal{I}_n) 
\end{equation*}
where $\mathcal{I}_n$ is the identity matrix of dimension $n \times n$. 

In this context, it is reasonable to choose the sample mean as summary statistic $\eta_0 = \frac{1}{n} \sum_{i=1}^n y_0^{(i)}$ and $\eta_{sim} = \frac{1}{n} \sum_{i=1}^n y_{sim}^{(i)}$, which is also sufficient for $\theta$, therefore this is a case where ABC targets the posterior distribution conditional on the observations and not only the summary statistics. In particular, the summary statistic $\eta$ follows a Gaussian distribution, $\eta_{sim} \sim \mathcal{N}(\theta, \frac{1}{n})$.

The synthetic likelihood is
\begin{equation}
\hat{L}_s(\theta) = \left( \frac{n}{2\pi} \right)^{1/2} \exp\left\{ -\frac{1}{2}(\eta_0-\theta-g)^2 \right\}
\label{eq:normal_slik}
\end{equation}
where $g$ is a random variable such that $g \sim \mathcal{N}(0, 1/(nN))$, and the estimator for $\theta$ follows again a Gaussian distribution: $\hat{\theta}_s \sim  \mathcal{N}(\eta_0, 1/(nN))$. 

\begin{figure}
    \centering
  \includegraphics[width=\textwidth, height=8cm]{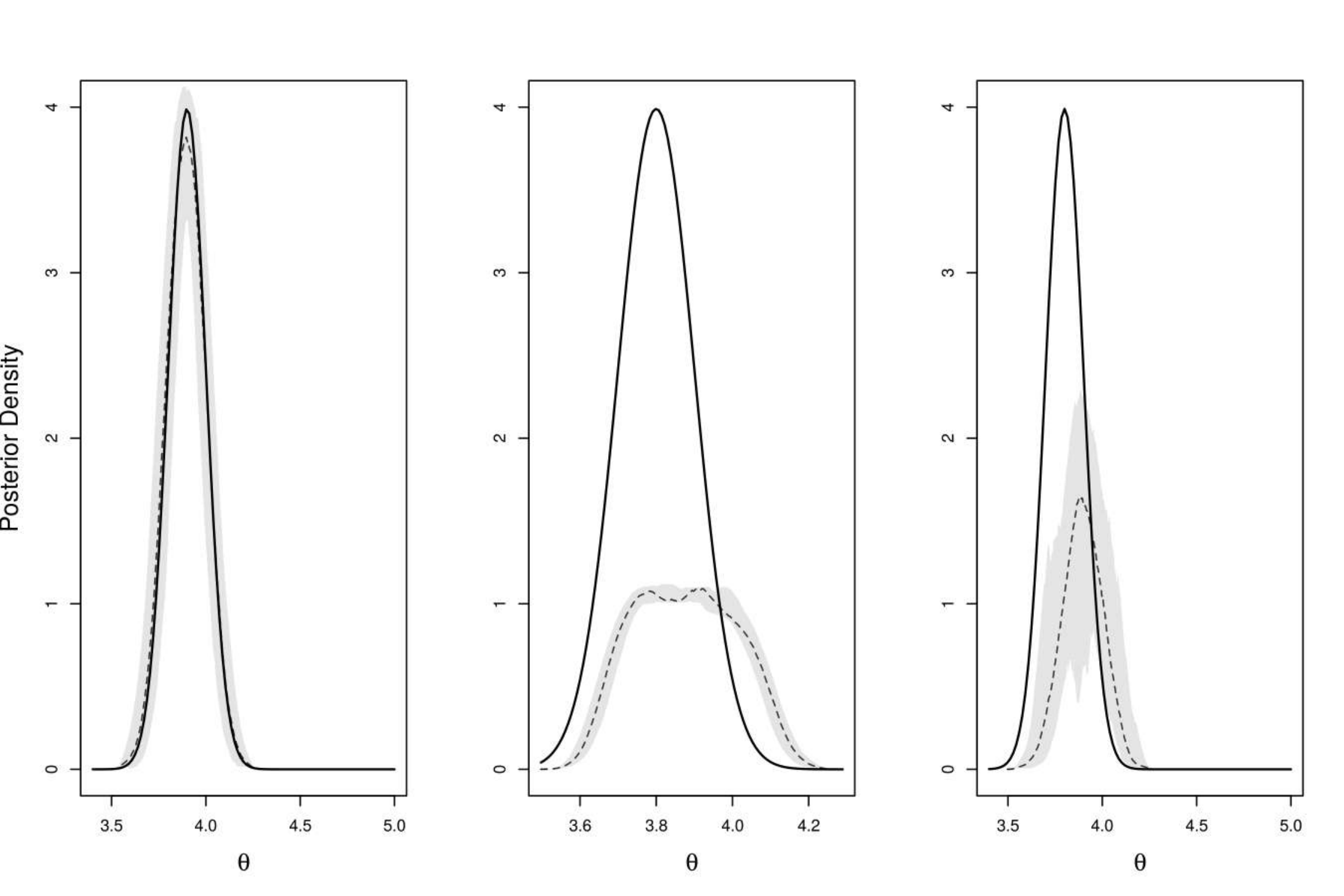}
  \caption{Approximations of the posterior distributions for $\theta$ based on synthetic likelihood (left), standard ABC (centre) and BOLFI (right) for the Gaussian example, with unknown location parameter and known variance. Solid line represents the true posterior distribution, dotted lines are the averages from 250 repetitions of the experiment, and the shaded area shows corresponding 95\% pointwise variability intervals.}
  \label{fig:normal_example}
\end{figure}

On the other hand, when using a nonparametric approximation, it is again reasonable to use the sample mean as the summary statistic in Algorithm \ref{alg1}. Suppose that $\Delta(\eta_0,\eta_{sim}) = (\eta_0-\eta_{sim})^2$. The probability that the discrepancy is lower than the threshold level $\varepsilon$ approximates the likelihood function:
\begin{align*}
\Pr \left(\Delta(\eta_0,\eta_{sim}) \leq \varepsilon \right) &= \Pr \left( (\eta_0-\eta_{sim})^2 \leq \varepsilon \right) \nonumber \\
&= \Pr \left( (\eta_0-\eta_{sim}) \geq \sqrt{\varepsilon} \right) + \Pr \left( (\eta_0-\eta_{sim}) \leq \sqrt{\varepsilon} \right) \nonumber \\
&= \Phi \left( \sqrt{n} (\eta_0 - \theta) + \sqrt{n\varepsilon} \right) - \Phi \left( \sqrt{n} (\eta_0 - \theta) - \sqrt{n\varepsilon} \right) 
\end{align*}
where $\Phi(\cdot)$ is the cumulative distribution function of a standard Gaussian variable. Therefore, when $n\varepsilon$ is small, the likelihood function of the approximate sample is
\begin{equation}
\hat{L}_{\varepsilon}(\theta) \propto \Pr \left(\Delta(\eta_0,\eta_{sim}) \leq \varepsilon \right) \propto \sqrt{\varepsilon} L(\theta),
\label{eq:normal_BOLFI}
\end{equation}
which means that the likelihood function of the parameter $\theta$ is well approximate for $\varepsilon$ small enough. However, for small $\varepsilon$ the acceptance rate is also small and a large amount of simulated values may be needed to obtain a good approximation of the posterior distribution of $\theta$. Figure \ref{fig:normal_example} shows the approximations of the posterior distribution of $\theta$ obtained with synthetic likelihood as described in Equation \eqref{eq:normal_slik}; standard ABC following Algorithm \ref{alg1} and the Bayesian optimization for likelihood-free inference (BOLFI) as in Equation \eqref{eq:normal_BOLFI}. The shaded area shows the 95\% variability intervals obtained by taking pointwise quantiles at the 0.025 and 0.975 level, over the 250 repeated simulations. 
In particular, the sample mean is chosen as the summary statistic for implementation of standard ABC and synthetic likelihood, moreover standard ABC is performed by keeping the Euclidean distance as discrepancy measure. The threshold $\varepsilon$ is chosen, here and in the following example, by performing a pilot run where the threshold is fixed at a relatively large value and then set at the 0.05 quantile of the empirical distribution of the distance among simulated and observed summary statistics in the pilot experiment. On the other hand, BOLFI is applied by using again the sample mean as the summary statistic, the Euclidean distance as discrepancy measure and a logarithmic transformation of the discrepancy to encourage normality. The first algorithm described in Section \ref{sec:nonparam} is performed, such that the discrepancy measure is assumed to follow a Gaussian distribution with constant variance. Moreover, the parameter $\theta$ is supposed to \textit{a priori} be defined between $[-20,20]$ and we chose $20$ initialization points sampled straight from the prior before starting to optimize the acquisition of samples; in order to save computational time, we decided to update the Gaussian process every 10 samples. All three algorithms have been applied to produce $1000$ samples of accepted parameter values. Computations were performed on a computer with a 2.5 GHz Intel Core i5 processor with 4GB 1600 Mhz DDr3 of RAM, the CPU times are: 9.42s  for synthetic likelihood (with 1000 summary statistics simulated), 1min 54s for standard ABC (with 20,000 simulated values), 1min 47s for BOLFI (4 chains of 1000 simulations with the first 75\% simulations used as burnin).

In this case, the synthetic likelihood produces the best approximation, closely following the true posterior distribution (solid lines), while the standard ABC approach was the least accurate, even the posterior mean here is not close to the truth, indicating the need to further reduce the tolerance $\varepsilon$. BOLFI represents an intermediate situation, slightly over-estimating the spread of the distributions. In this setting, a procedure based on an assumption of normality, as for the synthetic likelihood and BOLFI, increase the goodness of the approximation. As there is no loss of information in using the sample mean as summary statistics, the posterior distribution obtained by using the synthetic likelihood is approximating the true posterior distribution. The approximation obtained with BOLFI has an additional level of error due to the chosen threshold, but the normality assumption improves the approximation with respect to the standard ABC. Results can be improved by increasing the number of simulations.

\subsection{Ricker model}
\label{sub:ricker}

We now consider the well known Ricker model, introduced in Ecology by \shortciteA{ricker1954stock}. 
Suppose that the number of animals from a particular species is $\mathbf{y}_0$ and depends on the entire population which evolves dynamically. In particular, let
\begin{equation}
\log N^{(t)} = \log r + \log N^{(t-1)} - N^{(t-1)} + \sigma e^{(t)}
\label{eq:ricker_N}
\end{equation}
where $N^{(t)}$ is the unknown population at time $t$, $ \log r $ is the log-growth rate, $\sigma$ is the standard deviation of the innovation and $e^{(t)}$ are independent Gaussian errors. By assumption, $N^{(0)} = 0$. 

Then, the observed population at time $t$ is a Poisson random variable and the generative model can be described as 
\begin{equation}
y_{sim}^{(t)} \sim \mathcal{P}oi(\phi N^{(t)})
\label{eq:ricker_y}
\end{equation}
where $\phi$ is a scaling parameter. The likelihood function is difficult to evaluate, given the nonlinearity of the state equation and the fact that at every time $t$ it is necessary to integrate out the unobserved population $N_t$. 

\begin{figure}
    \centering
  \includegraphics[width=\textwidth]{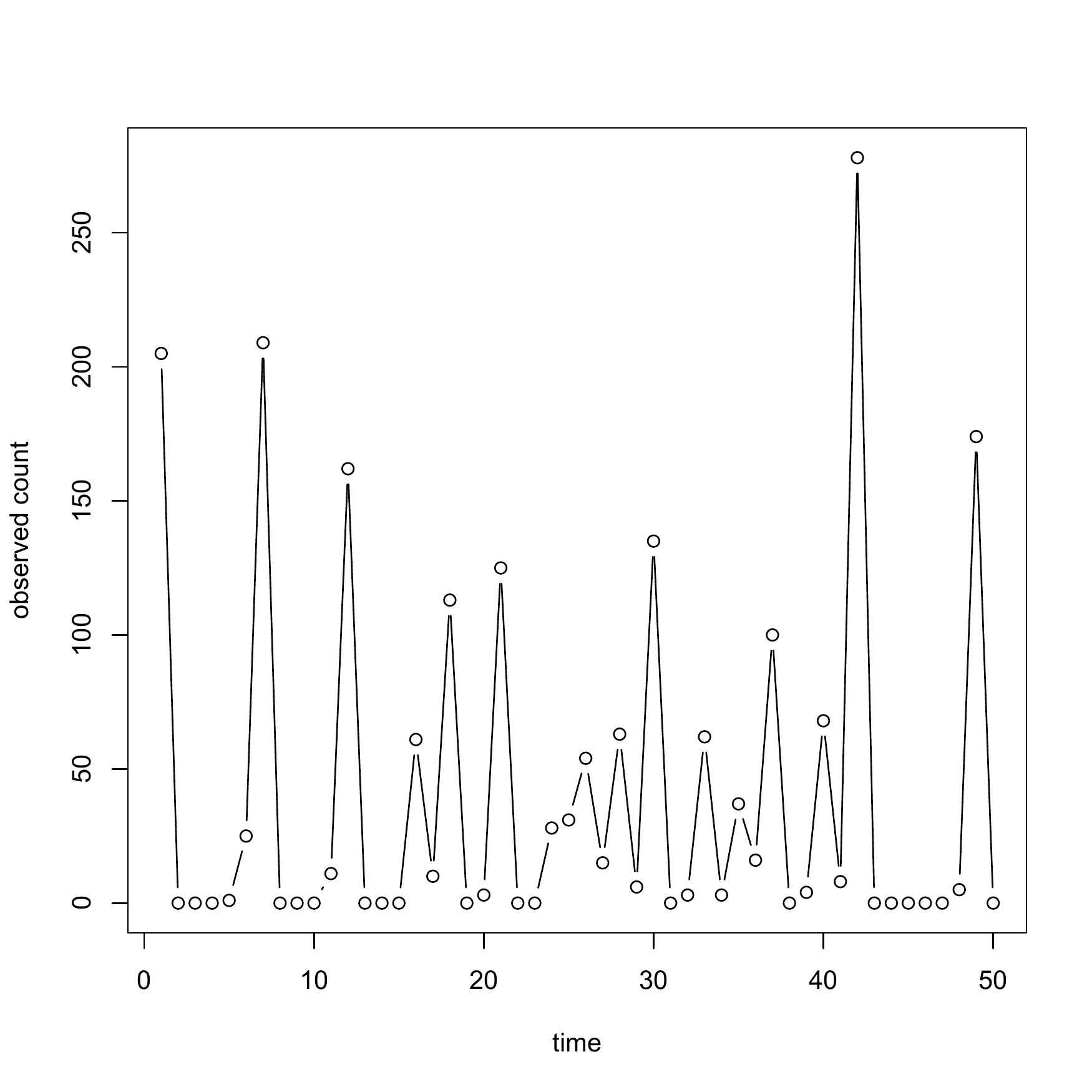}
  \caption{An example of simulated data from the Ricker model with $\log r = 4$, $\phi=10$ and $\sigma=0.3$.}
  \label{fig:ricker_obs}
\end{figure}

Following \shortciteA{gutmann2016bayesian}, we simulate data from model \eqref{eq:ricker_N} and \eqref{eq:ricker_y} by fixing $\sigma=0.3$, $\phi=10$ and $\log r = 4$. Inference is focused on the log-growth rate, while $\sigma$ and $\phi$ are considered nuisance parameters. Figure \ref{fig:ricker_obs} shows one example of dataset generated from the Ricker model with these choices of parameters and used to compare the algorithms. 

In order to apply the algorithms described in Section \ref{sec:syntlik} and Section \ref{sec:nonparam}, we 
perform inference with two sets of summary statistics: one set with five summary statistics, some of which can be considered (almost)-Gaussian distributed and a larger set of 13 summary statistics (proposed by \shortciteA{wood2010statistical}), whose distributions are less obviously Gaussian. The first set include: the number of observations greater than 10, the median count, the maximum count, the quantile of level $q=0.75$ and the sample mean of the observations greater than $1$. The second set include: the mean, the number of zeros, the
autocovariances up to lag 5 (including lag 0), the
parameter estimates of the regression model 
$$
y^{(t)^{0.3}} = \beta_0 y^{(t-1)^{0.3}} + \beta_1 y^{(t-1)^{0.6}} + \varepsilon_t
$$
with $\varepsilon_t \sim \mathcal{N}(0,\sigma^2)$, and the coefficients of a cubic regression of the ordered differences on their observed
values.

A threshold $\varepsilon = 5.0$ was chosen via a pilot run to standard ABC. The prior distributions for the parameters have been fixed as $\log r \sim \mathcal{U}(3,8)$, $\phi \sim \mathcal{U}(0,20)$ and $\sigma \sim \mathcal{U}(0,0.6)$. For each algorithm, we obtain again $1000$ samples of simulated and accepted parameter values. 
Using a computer with a 2.5 GHz Intel Core i5 processor with 4GB 1600 Mhz DDr3 of RAM, the CPU times for analysing one dataset when using 5 summary statistics are: 33s for synthetic likelihood (with 1000 summary statistics simulated), 0.0374s for standard ABC (with 6,000 simulated values), 1min 27s for BOLFI (with 4 chains of 1,000 simulations each and the first 75\% of the values used as burnin). When 13 summary statistics were used, the computing times were: 171s for synthetic likelihood (with 1000 summary statistics simulated), 4.58s for standard ABC (with 102,000 simulated values), 9min 31s for BOLFI (with 4 chains of 1,000 simulations each and the first 75\% of the values used as burnin).  For comparison, we have also implemented the same algorithms with a threshold $\varepsilon = 4.0$ and the computing CPU times are: 5.41s for standard ABC and 13min 50s for BOLFI.

\begin{figure}
    \centering
 \includegraphics[width=\textwidth, height=5cm]{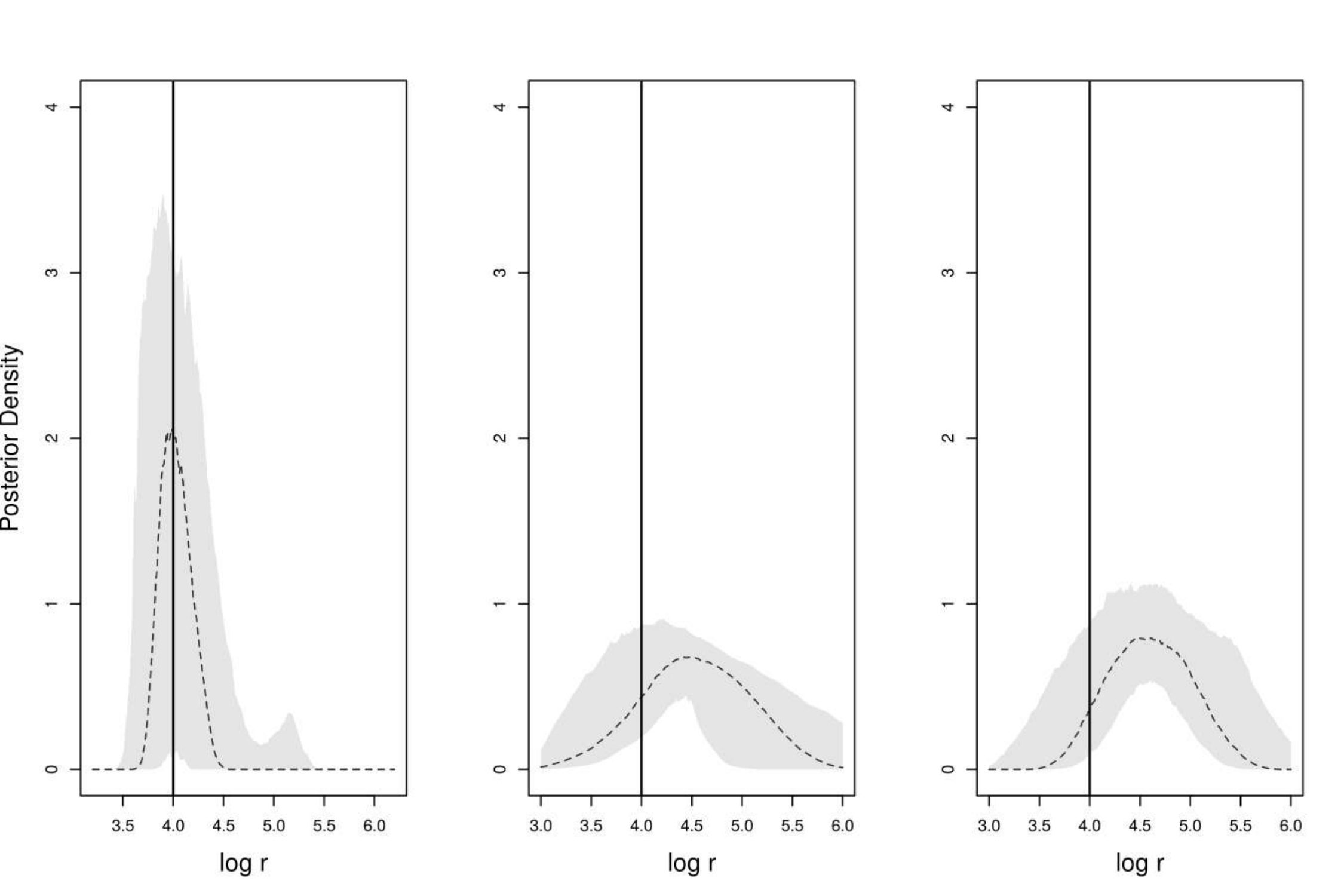}\\
 \includegraphics[width=\textwidth, height=5cm]{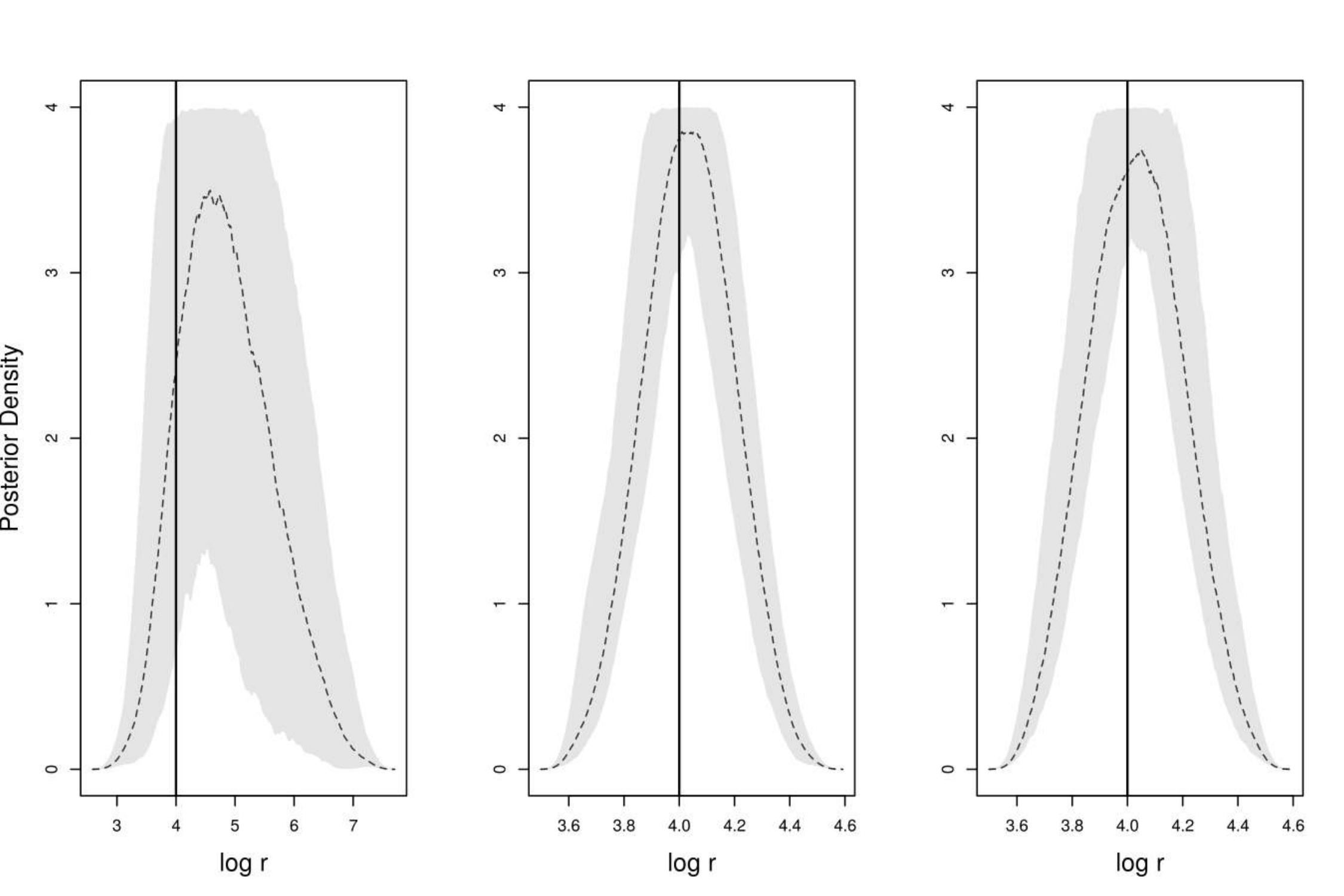}\\
  \caption{Approximations of the posterior distributions of $\log(r)$ based on synthetic likelihood (left), standard ABC (centre) and BOLFI (right) for the Ricker model example. Top row obtained from 5 summary statistics, and bottom row 13 summary statistics. Solid vertical lines represents the true value, dotted lines are the averages from
  250 repetitions of the experiment and the shaded areas show the 95\% pointwise variability intervals.}
    \label{fig:ricker_approx}
\end{figure}

In this case, the assumption of Gaussianity of the summary statistics is less justifiable. When using five summary statistics, the synthetic likelihood produced an approximation which is extremely concentrated around the true value from which the data have been generated; on the other hand, standard ABC and BOLFI are biased and  produced a large variability for the posterior, see Figure \ref{fig:ricker_approx} (top row). After increasing the number of summary statistics, the goodness of fit of the procedures changes: rejection ABC and BOLFI produced very similar approximations, concentrated around the true value of the parameter, while the approximation obtained with the synthetic likelihood seems to be less stable and became more uncertain and biased relative to the results from 5 summary statistics, suggesting a further deviation from normality in the extra summary statistics, see Figure \ref{fig:ricker_approx} (bottom row). It can be noticed that using improved versions of the standard ABC, like ABC-MCMC, can enhance the performance and reduce the uncertainty in the estimates, as shown in \shortciteA{fasiolo2016comparison}. On the other hand, using a larger number of summary statistics implies a greater computational effort, which is more evident for BOLFI and, partially, for ABC. 

 In general, increasing the number of summary statistics and reducing the tolerance level will help improve rejection ABC and BOLFI, while the synthetic likelihood approach is much more dependent on the normality
 assumptions on the summary statistics.
 It has to be noticed that in this example there is a large difference in computational times: one of the appealing characteristics of rejection ABC methods and methods based on the synthetic likelihood is that they are easily parallelizable. On the other hand, the sequential nature of BOLFI makes it computationally demanding, as it happens in this example. The advantage of BOLFI is that it relaxes the assumptions of the synthetic likelihood, while maintaining a parametric model. This can be useful in situations where standard ABC methods are extremely inefficient due to the simulation from a vague prior distribution or when it is difficult to identify regions of the parameter space of high posterior density. In this example, the chosen prior was in a region of high posterior density region, therefore the standard ABC algorithm was fast, however for many problems, it may not be possible or desirable to use an informative prior. In conclusion, we believe that BOLFI is an intermediate solution between the parametric approach of the synthetic likelihood and the standard ABC approach, as it is also evident from the results in Figure \ref{fig:normal_example}.

\section{Discussion}\label{sec:discussion}

There are many modifications designed to increase the computational efficiency of Algorithm \ref{alg1}. In general, simulating from areas of the parameter space where the likelihood function is not negligible increases the computational efficiency, as in the population Monte-Carlo approach of \shortciteA{marin2012approximate} or the sequential version of \shortciteA{sisson2007sequential}. See also \shortciteA{wilkinson2014accelerating}, \shortciteA{drovandi2018accelerating} and \shortciteA{jarvenpaa2019efficient} for efficient ways to simulate parameter values. 

Working with the summary statistics can also improve computational efficiency as well as estimation accuracy. For example, one can define a large number of summary statistics and choose the summary statistics in Algorithm \ref{alg1} as combinations of them (usually determined via regression); these approaches may be found in \shortciteA{nunes2010optimal}, \shortciteA{fearnhead2012constructing}, \shortciteA{
aeschbacher2012novel} and \shortciteA{blum2013comparative}.

An active area of research not covered in this review focuses on the choice of discrepancy measure based on classification accuracy, as the need for comparing simulated and observed datasets may be seen as a classification problem, and this creates a natural connection with approaches developed in machine learning, see for example \shortciteA{gutmann2018likelihood}.  The basic idea is that it should be straightforward to discriminate between datasets which have been generated by very different values of the parameters.
In this case, the choice of the summary statistics is replaced by the choice of a classification method. However, while the use of summary statistics may lead to a loss of information, the classification rule is learned from the data, therefore its choice has a smaller impact on the quality of the approximation.  
While a thorough analysis and comparison of classification methods has not been performed yet, it is likely that the performance of a classification method will be problem dependent.

Finally, we have not implemented the methods in Sections \ref{sec:machine}, but we feel this direction could take ABC to a higher level: to the realm of big data and high dimensions and less user-specific tuning.

\section*{Acknowledgements}

The authors thank Michael Gutmann and Henri Personen for useful conversations. In particular, the authors thank Henri Personen for his help with the code used for BOLFI in the examples. 

The work of the first author has been partially developed under the PRIN2015 supported-project: ``Environmental processes and human
activities: capturing their interactions via statistical methods (EPHAStat)'', funded by MIUR (Italian Ministry of Education, University and
Scientific Research) (20154X8K23-SH3). The second author is grateful for the support of the Australian Reearch Council Center of Excellence for Mathematical and Statistial Frontiers.

\end{document}